\documentclass[twocolumn,prb,aps,floatfix,nobalancelastpage,amsfonts,citeautoscript]{revtex4}

\usepackage{graphicx}
\usepackage{bm}

\begin{document}

\title{Fermi surface pockets in ortho-II YBa$_2$Cu$_3$O$_{6.5}$: the origin of quantum oscillations?}
\author{A. Carrington, and E.A. Yelland }
\affiliation{H.H. Wills Physics Laboratory, University of Bristol, Tyndall Avenue, Bristol, United
Kingdom.}

\date{\today}

\begin{abstract}
In this paper we explore whether the quantum oscillation signals recently observed in ortho-II
YBa$_2$Cu$_3$O$_{6.5}$ (OII-Y123) may be explained by conventional density functional
band-structure theory. Our calculations show that the Fermi surface of OII-Y123 is extremely
sensitive to small shifts in the relative positions of the bands. With rigid band shifts of around
$\pm$30 meV small tubular pockets of Fermi surface develop around the Y point in the Brillouin
zone. The cross-sectional areas and band masses of the quantum oscillatory orbits on these pockets
are close to those observed.  The differences between the band-structure of OII-Y123 and
YBa$_2$Cu$_4$O$_{8}$ (Y124) are discussed with reference to the very recent observation of quantum
oscillations in Y124.
\end{abstract}
\maketitle

The nature of the normal state of the high temperature cuprate superconductors (HTC) has been a
topic of intense discussion ever since their discovery.  The unusual temperature dependence shown
by the resistivity and Hall coefficient for example, and how these evolve as a function of doping,
has led to a wide range of exotic theories of the nature of the normal state \cite{hussey07}. Many
of these theories depart substantially from the standard Fermi liquid model of a metal,
particularly in the underdoped region of the phase diagram.

The recent report \cite{Doiron-leyraudPLLBLBHT07} of the observation of Shubnikov-de Haas (SdH)
oscillations in the Hall and longitudinal resistivities of underdoped ortho-II
YBa$_2$Cu$_3$O$_{6.5}$ (OII-Y123) was a surprising and potentially extremely important result.
These oscillations suggest that even in the underdoped region some well-defined pockets of Fermi
surface exist.  The observed orbits have a frequency of $\sim 530\pm20$~T and a mass,
$m^*=1.9\pm0.1 m_e$ ($m_e$ is the free electron mass). A single pocket with this dHvA frequency
corresponds to a small ($\sim$ 2\%) fraction of the Brillouin zone. Identifying the origin of this
orbit is clearly very important for its interpretation.  In this paper, we discuss whether the SdH
signals could come from small pockets of Fermi surface predicted by conventional density functional
band-structure calculations.

\begin{figure*}
\includegraphics*[width=0.98\linewidth,clip]{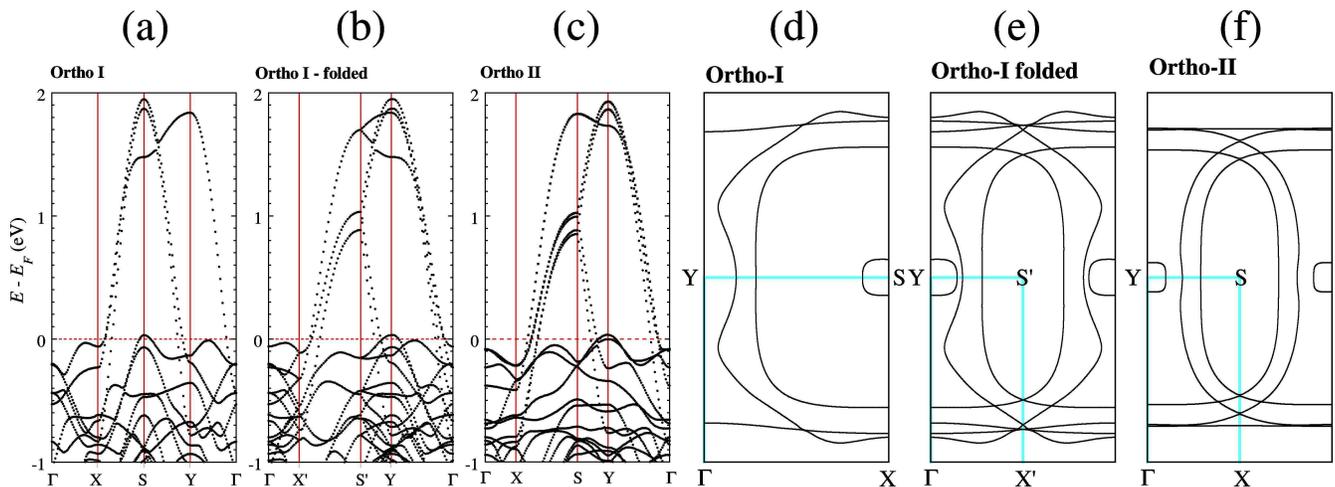}
\caption{(Color online) Panels (a-c): The band-structure of YBa$_2$Cu$_3$O$_{7-\delta}$. (a)
OI-Y123, (b) OI-Y123 folded in to the reduced zone of OII-Y123 (c) OII-Y123. Panels (d-f) Fermi
surfaces of YBa$_2$Cu$_3$O$_{7-\delta}$. (d) OI-Y123, (e) OI-Y123 folded, (f) OII-Y123 (with
$\Delta E_F$=+20~meV).} \label{figspag}
\end{figure*}

Our understanding of the electronic structure of the cuprates has, to date, mostly been led by
angle resolved photoemission spectroscopy (ARPES). The resolution of this technique has advanced
rapidly over the past decade and had provided many key insights into the physics of HTC
\cite{DamascelliHS03}.   More recently, the strong angle dependent magnetoresistance of strongly
overdoped Tl$_2$Ba$_2$CuO$_{6+\delta}$ has been used to extract information about the Fermi surface
and scattering rate \cite{HusseyACMB03,Abdel-jawadKBCMMH06}.  The shape and size of the Fermi
surfaces measured by both these techniques are generally in very good agreement with conventional
density functional theory (DFT) band-structure calculations \cite{DamascelliHS03,Sahrakorp07}.

Quantum oscillatory effects, e.g.\ the de Haas-van Alphen (dHvA) and Shubnikov-de Haas effects, are
very powerful probes of the Fermi surface properties of a metal.  Unlike ARPES, they probe the bulk
of the material (and so are not sensitive to surface defects) and are a true three dimensional
probe of the quasiparticles at the Fermi level. The analysis of the amplitude and frequency of the
signals and how these change as a function of magnetic field and temperature, gives information
about the Fermi surface and the quasiparticle masses and scattering rates. The main constraint is
that impurities severely attenuate the SdH/dHvA signal, according to the Dingle factor
$R_D=\exp(-\frac{\pi \hbar k_F}{ eB\ell})$ (here $\pi k_F^2=\mathcal{A}$ the cross-sectional area
of the dHvA orbit and $\ell$ is the orbitally averaged mean free path). The small size of the orbit
in OII-Y123 was a key factor in it being observable.

Band-structure calculations of fully oxygenated ortho-I YBa$_2$Cu$_3$O$_{7}$ (OI-Y123) show a small
tubular pocket of Fermi surface near the S point which derives from the `CuO chain bands'
\cite{PickettCK90,Pickett89,AndersenLJP95} (see below). This could be a simple explanation for the
origin of the observed SdH orbit.  However the band-structure calculations of Bascones \textit{et
al.} \cite{BasconesRSLA05} indicate that this small pocket is absent in OII-Y123. Another
intriguing possibility is that the SdH signal comes from pockets close to the nodal regions, which
may have formed because of Fermi surface reconstruction. In this paper we calculate the
band-structure of OII-Y123 using conventional density functional theory techniques and investigate
the possible extremal orbits which could give rise to the observed orbit.  We contrast this with
similar calculations on the double-chain cuprate YBa$_2$Cu$_4$O$_{8}$ (Y124) in which SdH
oscillations have also recently been observed \cite{Yelland07,Bangura07}.

Our calculations were carried out using the Wien2K package \cite{wien2k}, which is an
implementation of a full-potential, augmented plane wave plus local orbital scheme.  We used a
generalized gradient approximation form for the exchange correlation potential
\cite{PerdewBE96,lsdanote}, and the crystal structure of YBa$_2$Cu$_3$O$_{6.5}$ determined by
Grybos \textit{et al.} \cite{GrybosHZSKEW94,GrybosWGT01}. A dense $k$-mesh of $\sim 10^4$ points in
the full Brillouin zone (19$\times$39$\times$12, or 1400 points in the irreducible wedge of the
Brillouin zone) was used for the self-consistency cycle.

The calculated band-structure in the basal plane at the center of the Brillouin zone ($k_z=0$) is
shown in Fig.\ \ref{figspag}. The main features can be easily related to that of fully oxygenated
OI-Y123 by band folding. In the first panel of the Fig.\ \ref{figspag}(a) we show our calculation
of the band-structure of OI-Y123 which is almost identical to previous calculations (e.g., Refs.\
\onlinecite{PickettCK90,AndersenLJP95}). The OI results have then been folded down into the smaller
OII Brillouin zone produced by doubling the $a$ lattice parameter (panel b). It can been seen that
this procedure reproduces many of the features of the full OII-Y123 calculation (panel c). The
effect of the band folding on the Fermi surface is also shown in the figure and again compared to
the results from the OII-Y123 calculation (panels d-f).

Moving along the X-S line the first band to cross the Fermi level is mainly due to the CuO chain
(Fig.\ \ref{figspag}c). It can be seen that this is very close to being one-dimensional and has
much less dispersion compared to the OI result. Note that there is only one chain band crossing in
the OII calculation compared to two in the folded OI calculation because OII only has one
conducting CuO chain per unit cell. The next four crossings are due to the CuO$_2$ planes. As in
the OI calculation there is a sizeable splitting between the bonding and antibonding CuO$_2$ bands
($\sim$ 210 meV along the X-S line at $E_F$). Each of these two bands are further split into two by
the additional $2a$ periodicity.  This is most evident close to the S point. We find that this
splitting is very small at $E_F$ along the X-S line, being $\sim 17\pm2$ meV. In the OI-Y123
calculation a fairly flat CuO/BaO band, arising mainly from hopping between chain oxygen sites via
the apical O site in the BaO layer below it, passes through $E_F$ close to the S point and gives
rise to a small tubular (quasi-two-dimensional) hole pocket
\cite{Pickett89,MazinJALRU92,AndersenLJP95}. Band folding moves this same pocket to the Y point in
the small (OII) zone. In the full OII-Y123 calculation, this band also crosses $E_F $ and again
forms a tubular hole pocket close to the Y point. In addition, a second band with the same
character is very close in energy, $\sim 35$~meV lower at Y.

Our results are similar to a calculation of OII-Y123 reported by Bascones \textit{et al.} \cite{BasconesRSLA05}  The
main differences are that the splittings of the CuO$_2$ bands due to the $2a$ periodicity are larger and near to Y the
CuO/BaO bands are more closely spaced (and $\sim$ 50 meV lower in energy) than in our calculation.  Bascones \textit{et
al.} used a TB-LMTO-ASA method which may be less accurate for a non-close packed structure like OII-Y123 than the full
potential method used here. In any case, these differences would not change our essential conclusions.

The shape and size of the two plane sheets predicted for OI-Y123 have been verified by ARPES
measurements on YBa$_2$Cu$_3$O$_{6.95}$ \cite{SchabelMPMSBLRH98}.  There is also some weak evidence
for the quasi-1D chain sheet but there is no sign of the hole pocket centered on S. However, it
should be mentioned that ARPES measurements on YBCO are complicated by a surface feature
originating from the CuO chains, which may limit the resolution for Fermi surface sheets with
significant chain character \cite{SchabelMPMSBLRH98}.

\begin{figure*}[t]
\center
\includegraphics*[width=0.98\linewidth,clip]{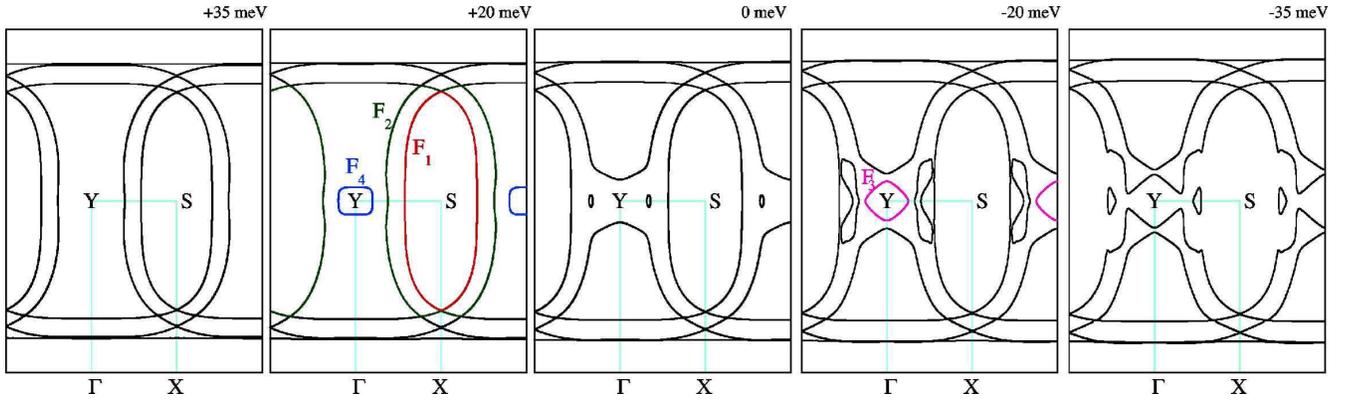}
\caption{(Color online) Evolution of the Fermi surface of OII-YBa$_2$Cu$_3$O$_{6.5}$ with Fermi
level shift $\Delta E_F$. The figure shows two-dimensional cuts within the basal plane ($k_z=0$),
as indicated by the symmetry labels. The primary quantum oscillation orbits ($F_n$) are marked on
the +20 meV and -20 meV panels.} \label{figEfscan}
\end{figure*}

Although DFT calculations usually correctly describe the general features of the band-structure,
 there are often small discrepancies with respect to the relative positions of the bands when
compared to experiment. This is true even if the calculation has been converged to the meV level. Sr$_2$RuO$_4$ is an
example of an oxide material, with a similar structure to the cuprate superconductor La$_{2-x}$Sr${_x}$CuO$_{4}$, which
has been studied in detail by dHvA measurements. To get agreement with band-structure calculations the bands need to be
shifted by $\sim$40~meV, in opposite directions \cite{MackenzieJDMRLMNF96}. Even in the relatively simple material
MgB$_2$, shifts of order 100~meV are needed \cite{CarringtonMCBHYLYTKK03,CarringtonYFC07}.

\begin{figure}[b]
\includegraphics*[width=0.98\linewidth,clip]{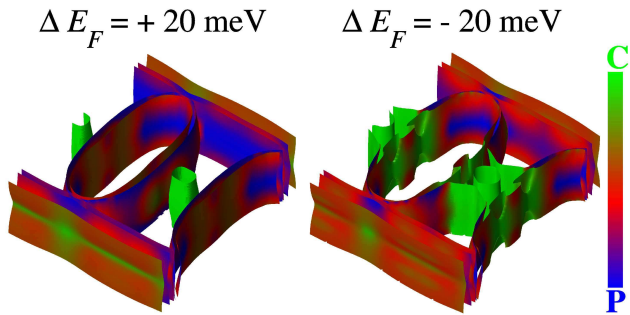}
\caption{(Color online) Three dimensional image of the calculated Fermi surface of OII-Y123 showing
the small pockets centered on the Y points for the two different Fermi level shifts indicated. The
shading indicates the band character C=Chain like (including apical oxygen), P=plane like.}
\label{fig3dfs}
\end{figure}

In Fig.\ \ref{figEfscan} we show how the Fermi surface of OII-Y123 changes as the $E_F$ is varied by $\pm$35 meV. These
small rigid band shifts correspond to adding (removing) 0.05 (0.07) electrons per Cu atom, although a similar result
could be obtained without doping by moving the plane and chain bands in opposite directions. For $\Delta E_F$=+35meV
the Fermi surface consists of two large hole-like tubular CuO$_2$ sheets centered on S, plus three
quasi-one-dimensional sheets (one from the chains and two from the planes).  As $E_F$ is reduced a small hole-like
pocket develops near the Y point, the origin of which (as discussed above) is the same as that of the pocket near the S
point in OI-Y123. Further reduction of $E_F$ results in this pocket growing in size and then merges with the
antibonding CuO$_2$ plane sheet.  As $E_F$ is further reduced the second CuO/BaO band passes through the Fermi level
giving rise to another pocket.  Eventually this merges with the bonding CuO$_2$ plane sheet.  Both of the pockets that
open close to Y are quasi-two-dimensional (tubular) sheets, with relatively weak warping along the \textit{c} direction
(see Fig.\ \ref{fig3dfs}).  Similar results to these have recently been report by Elfimov \textit{et al.}
\cite{ElfimovSD07}.

\begin{figure}
\includegraphics*[width=0.98\linewidth,clip]{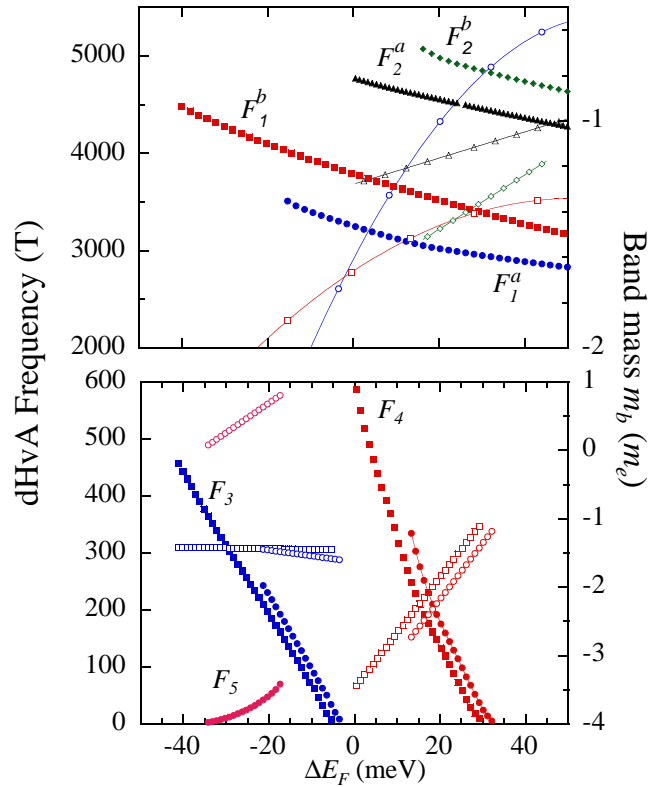}
\caption{(Color online) Extremal dHvA frequencies versus Fermi level shift (solid symbols). The
corresponding band masses are shown with open symbols (right-hand axis).  } \label{figdHvAfreq}
\end{figure}

\begin{figure}
\includegraphics*[width=0.5\linewidth,clip]{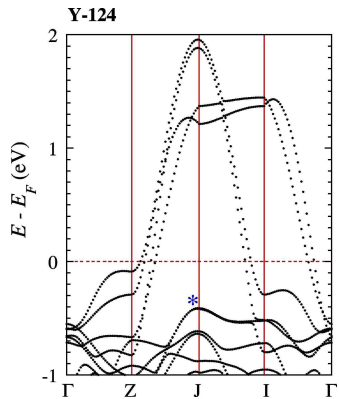}
\caption{(Color online) Band-structure of YBa$_2$Cu$_4$O$_{8}$ (Y124).  The symmetry labels
appropriate for the Ammm space group of Y124 are the same as in Ref.\
\onlinecite{AmbroschdraxlBS91}. The path $\Gamma$ZJI is approximately equivalent to the $\Gamma$XSY
path used in the other figures for the Pmmm space group of OI/II-Y123.} \label{figY124Spag}
\end{figure}

The above illustrates the extreme sensitivity of the Fermi surface to the relative positions of the
bands and means that subtle changes to the doping could result in small Fermi surface pockets being
formed.  It is possible that one such pocket could give rise to the quantum oscillation signals
observed by Doiron-Leyraud \textit{et al.} \cite{Doiron-leyraudPLLBLBHT07}. To investigate this
possibility more fully we have calculated the quantum oscillation frequencies [$F=(\hbar /2\pi
e)\mathcal{A} $] and band masses [$m_b=\frac{\hbar^2}{2\pi}\frac{\partial \mathcal{A}}{\partial
E}$]. The frequencies shown in the bottom panel of Fig.\ \ref{figdHvAfreq} ($F_3$ and $F_4$) are
from the small hole pockets discussed above, whereas those in the upper panel ($F_1$ and $F_2$) are
from the main CuO$_2$ sheet surfaces.  Each of the tubular sections of Fermi surface has a minimum
and maximum extremal area which have been labeled $F_i^a$ and  $F_i^b$ respectively in the figure.
As expected, the extremal areas (dHvA frequencies) of both hole pockets vary strongly with the
Fermi level shift and have maximum frequencies of 400-600~T. The band mass of the pocket $F_3$ is
roughly constant as a function of $\Delta E_F$ with an average value of $\sim$ 1.5 $m_e$.  The mass
of the other pocket orbit, $F_4$ is much more variable, ranging from $\sim 1.5$ for large $\Delta
E_F$ to $\sim 3$ as $\Delta E_F$ approaches zero.  A similar calculation for OI-Y123 was reported
in Ref.\ \onlinecite{MazinJALRU92}.

The experimentally observed orbit frequency ($F=530\pm20$~T) is at the upper limit of the
calculated range for both $F_3$ and $F_4$.  The observed mass of the orbit ($m^*=1.9\pm0.1m_e$)
seems to be more compatible with $F_3$ than $F_4$.  We expect that the observed  quasiparticle mass
should be enhanced compared to the band mass [$m^*=(1+\lambda)m_b$] because of many body
interactions (electron-phonon and electron-electron) which are not included in the present
calculations. Although it is expected that $\lambda$ will be relatively large ($\sim 1-2$) on the
CuO$_2$ planes \cite{LanzaraBZKFLYEFKSNUHS01} it could be substantially less on these sheets of
Fermi surface without significant plane character. Hence, the calculated mass of $F_3$ could be
consistent with experiment.  Experimental studies of the pressure or doping dependence of the SdH
frequencies in Y123 may help determine this interpretation is correct. An alternative approach is
to study other cuprate materials where the CuO/BaO band is either absent or is at markedly
different energies.

Very recently two groups \cite{Yelland07,Bangura07} have reported the observation of SdH
oscillation in Y124 with a comparable frequency to those in OII-Y123.  Y124 is an intrinsically
underdoped cuprate with a similar structure to OI-Y123, but with a double chain layer and a doubled
unit cell along the $c$-direction. Our calculations of the band-structure, which are essentially
the same as reported previously \cite{AmbroschdraxlBS91,YuPF91}, show that the CuO/BaO band which
gives rise to the pockets in Y123 is now around 400~meV below $E_F$ at the zone corner J (marked
with an asterisk in Fig.\ \ref{figY124Spag}). Hence, it is very unlikely that this band could be
responsible for the experimental observations. In fact, we not find any orbits within $\Delta
E_F=\pm400$ meV which have frequencies close to experiment.

In summary, we have presented calculations of the electronic structure of OII-Y123, and have
discussed possible origins of low frequency quantum oscillation signals.  Two possible orbits, both
arising from  chain/apical oxygen bands have been identified, which with a small shift in the
relative energies of the bands, have extremal areas compatible with the observed SdH signals. The
band masses of the quasiparticles on at least one of these orbits is roughly consistent with those
observed.  It seems unlikely however that the reported orbits in Y124 can be explained by a similar
band-structure approach.

We thank S.B.~Dugdale, N.E.~Hussey, and S.M.~Hayden for useful comments.


\begin{thebibliography}{26}
\expandafter\ifx\csname natexlab\endcsname\relax\def\natexlab#1{#1}\fi \expandafter\ifx\csname
bibnamefont\endcsname\relax
  \def\bibnamefont#1{#1}\fi
\expandafter\ifx\csname bibfnamefont\endcsname\relax
  \def\bibfnamefont#1{#1}\fi
\expandafter\ifx\csname citenamefont\endcsname\relax
  \def\citenamefont#1{#1}\fi
\expandafter\ifx\csname url\endcsname\relax
  \def\url#1{\texttt{#1}}\fi
\expandafter\ifx\csname urlprefix\endcsname\relax\def\urlprefix{URL }\fi
\providecommand{\bibinfo}[2]{#2} \providecommand{\eprint}[2][]{\url{#2}}

\bibitem[{\citenamefont{Hussey}(2007)}]{hussey07}
\bibinfo{author}{\bibfnamefont{N.~E.} \bibnamefont{Hussey}},
  \emph{\bibinfo{title}{Normal State Transport Properties}}
  (\bibinfo{publisher}{Springer - Holland}, \bibinfo{year}{2007}),
  \bibinfo{note}{in `Handbook of High Temperature Superconductivity: Theory and
  Experiment', by J. Brooks (Adapter), J. R. Schrieffer (Editor)}.

\bibitem[{\citenamefont{Doiron-Leyraud
  et~al.}(2007)\citenamefont{Doiron-Leyraud, Proust, Leboeuf, Levallois,
  Bonnemaison, Liang, Bonn, Hardy, and Taillefer}}]{Doiron-leyraudPLLBLBHT07}
\bibinfo{author}{\bibfnamefont{N.}~\bibnamefont{Doiron-Leyraud}},
  \bibinfo{author}{\bibfnamefont{C.}~\bibnamefont{Proust}},
  \bibinfo{author}{\bibfnamefont{D.}~\bibnamefont{Leboeuf}},
  \bibinfo{author}{\bibfnamefont{J.}~\bibnamefont{Levallois}},
  \bibinfo{author}{\bibfnamefont{J.~B.} \bibnamefont{Bonnemaison}},
  \bibinfo{author}{\bibfnamefont{R.~X.} \bibnamefont{Liang}},
  \bibinfo{author}{\bibfnamefont{D.~A.} \bibnamefont{Bonn}},
  \bibinfo{author}{\bibfnamefont{W.~N.} \bibnamefont{Hardy}}, \bibnamefont{and}
  \bibinfo{author}{\bibfnamefont{L.}~\bibnamefont{Taillefer}},
  \bibinfo{journal}{Nature} \textbf{\bibinfo{volume}{447}},
  \bibinfo{pages}{565} (\bibinfo{year}{2007}).

\bibitem[{\citenamefont{Damascelli et~al.}(2003)\citenamefont{Damascelli,
  Hussain, and Shen}}]{DamascelliHS03}
\bibinfo{author}{\bibfnamefont{A.}~\bibnamefont{Damascelli}},
  \bibinfo{author}{\bibfnamefont{Z.}~\bibnamefont{Hussain}}, \bibnamefont{and}
  \bibinfo{author}{\bibfnamefont{Z.~X.} \bibnamefont{Shen}},
  \bibinfo{journal}{Rev. Mod. Phys.} \textbf{\bibinfo{volume}{75}},
  \bibinfo{pages}{473} (\bibinfo{year}{2003}).

\bibitem[{\citenamefont{Hussey et~al.}(2003)\citenamefont{Hussey, Abdel-jawad,
  Carrington, Mackenzie, and Balicas}}]{HusseyACMB03}
\bibinfo{author}{\bibfnamefont{N.~E.} \bibnamefont{Hussey}},
  \bibinfo{author}{\bibfnamefont{M.}~\bibnamefont{Abdel-jawad}},
  \bibinfo{author}{\bibfnamefont{A.}~\bibnamefont{Carrington}},
  \bibinfo{author}{\bibfnamefont{A.~P.} \bibnamefont{Mackenzie}},
  \bibnamefont{and} \bibinfo{author}{\bibfnamefont{L.}~\bibnamefont{Balicas}},
  \bibinfo{journal}{Nature} \textbf{\bibinfo{volume}{425}},
  \bibinfo{pages}{814} (\bibinfo{year}{2003}).

\bibitem[{\citenamefont{Abdel-jawad et~al.}(2006)\citenamefont{Abdel-jawad,
  Kennett, Balicas, Carrington, Mackenzie, Mckenzie, and
  Hussey}}]{Abdel-jawadKBCMMH06}
\bibinfo{author}{\bibfnamefont{M.}~\bibnamefont{Abdel-jawad}},
  \bibinfo{author}{\bibfnamefont{M.~P.} \bibnamefont{Kennett}},
  \bibinfo{author}{\bibfnamefont{L.}~\bibnamefont{Balicas}},
  \bibinfo{author}{\bibfnamefont{A.}~\bibnamefont{Carrington}},
  \bibinfo{author}{\bibfnamefont{A.~P.} \bibnamefont{Mackenzie}},
  \bibinfo{author}{\bibfnamefont{R.~H.} \bibnamefont{Mckenzie}},
  \bibnamefont{and} \bibinfo{author}{\bibfnamefont{N.~E.}
  \bibnamefont{Hussey}}, \bibinfo{journal}{Nat. Phys.}
  \textbf{\bibinfo{volume}{2}}, \bibinfo{pages}{821} (\bibinfo{year}{2006}).

\bibitem[{\citenamefont{S.~Sahrakorpi et~al.}(2006)\citenamefont{S.~Sahrakorpi,
  Lin, Markiewicz, and Bansil}}]{Sahrakorp07}
\bibinfo{author}{\bibfnamefont{S.}~\bibnamefont{S.~Sahrakorpi}},
  \bibinfo{author}{\bibfnamefont{H.}~\bibnamefont{Lin}},
  \bibinfo{author}{\bibfnamefont{R.}~\bibnamefont{Markiewicz}},
  \bibnamefont{and} \bibinfo{author}{\bibfnamefont{A.}~\bibnamefont{Bansil}},
  \bibinfo{journal}{cond-mat/0607132}  (\bibinfo{year}{2006}).

\bibitem[{\citenamefont{Pickett et~al.}(1990)\citenamefont{Pickett, Cohen, and
  Krakauer}}]{PickettCK90}
\bibinfo{author}{\bibfnamefont{W.~E.} \bibnamefont{Pickett}},
  \bibinfo{author}{\bibfnamefont{R.~E.} \bibnamefont{Cohen}}, \bibnamefont{and}
  \bibinfo{author}{\bibfnamefont{H.}~\bibnamefont{Krakauer}},
  \bibinfo{journal}{Phys. Rev. B} \textbf{\bibinfo{volume}{42}},
  \bibinfo{pages}{8764} (\bibinfo{year}{1990}).

\bibitem[{\citenamefont{Pickett}(1989)}]{Pickett89}
\bibinfo{author}{\bibfnamefont{W.~E.} \bibnamefont{Pickett}},
  \bibinfo{journal}{Rev. Mod. Phys.} \textbf{\bibinfo{volume}{61}},
  \bibinfo{pages}{433} (\bibinfo{year}{1989}).

\bibitem[{\citenamefont{Andersen et~al.}(1995)\citenamefont{Andersen,
  Liechtenstein, Jepsen, and Paulsen}}]{AndersenLJP95}
\bibinfo{author}{\bibfnamefont{O.~K.} \bibnamefont{Andersen}},
  \bibinfo{author}{\bibfnamefont{A.~I.} \bibnamefont{Liechtenstein}},
  \bibinfo{author}{\bibfnamefont{O.}~\bibnamefont{Jepsen}}, \bibnamefont{and}
  \bibinfo{author}{\bibfnamefont{F.}~\bibnamefont{Paulsen}},
  \bibinfo{journal}{J. Phys. Chem. Solids} \textbf{\bibinfo{volume}{56}},
  \bibinfo{pages}{1573} (\bibinfo{year}{1995}).

\bibitem[{\citenamefont{Bascones et~al.}(2005)\citenamefont{Bascones, Rice,
  Shorikov, Lukoyanov, and Anisimov}}]{BasconesRSLA05}
\bibinfo{author}{\bibfnamefont{E.}~\bibnamefont{Bascones}},
  \bibinfo{author}{\bibfnamefont{T.~M.} \bibnamefont{Rice}},
  \bibinfo{author}{\bibfnamefont{A.~O.} \bibnamefont{Shorikov}},
  \bibinfo{author}{\bibfnamefont{A.~V.} \bibnamefont{Lukoyanov}},
  \bibnamefont{and} \bibinfo{author}{\bibfnamefont{V.~I.}
  \bibnamefont{Anisimov}}, \bibinfo{journal}{Phys. Rev. B}
  \textbf{\bibinfo{volume}{71}}, \bibinfo{pages}{012505}
  (\bibinfo{year}{2005}).

\bibitem[{Yel()}]{Yelland07}
\bibinfo{note}{E. A. Yelland, J. Singleton, C. H. Mielke, N. Harrison, F. F.
  Balakirev, B. Dabrowski, and J. R. Cooper, arXiv:0707.0057v1
  [cond-mat.supr-con]}.

\bibitem[{Ban()}]{Bangura07}
\bibinfo{note}{A. F. Bangura, J. D. Fletcher, A. Carrington, J. Levallois, M.
  Nardone, B. Vignolle, P. J. Heard, N. Doiron-Leyraud, D. LeBoeuf, L.
  Taillefer, S. Adachi, C. Proust, and N. E. Hussey, arXiv:0707.4461v1
  [cond-mat.supr-con]}.

\bibitem[{\citenamefont{Blaha et~al.}(2001)\citenamefont{Blaha, Schwarz,
  Madsen, Kvasnicka, and Luitz}}]{wien2k}
\bibinfo{author}{\bibfnamefont{P.}~\bibnamefont{Blaha}},
  \bibinfo{author}{\bibfnamefont{K.}~\bibnamefont{Schwarz}},
  \bibinfo{author}{\bibfnamefont{G.~K.~H.} \bibnamefont{Madsen}},
  \bibinfo{author}{\bibfnamefont{D.}~\bibnamefont{Kvasnicka}},
  \bibnamefont{and} \bibinfo{author}{\bibfnamefont{J.}~\bibnamefont{Luitz}},
  \emph{\bibinfo{title}{WIEN2K, An Augmented Plane Wave + Local Orbitals
  Program for Calculating Crystal Properties}} (\bibinfo{publisher}{Karlheinz
  Schwarz, Techn. Universit\"at Wien, Austria}, \bibinfo{year}{2001}),
  \bibinfo{note}{ISBN 3-9501031-1-2}.

\bibitem[{\citenamefont{Perdew et~al.}(1996)\citenamefont{Perdew, Burke, and
  Ernzerhof}}]{PerdewBE96}
\bibinfo{author}{\bibfnamefont{J.~P.} \bibnamefont{Perdew}},
  \bibinfo{author}{\bibfnamefont{K.}~\bibnamefont{Burke}}, \bibnamefont{and}
  \bibinfo{author}{\bibfnamefont{M.}~\bibnamefont{Ernzerhof}},
  \bibinfo{journal}{Phys. Rev. Lett.} \textbf{\bibinfo{volume}{77}},
  \bibinfo{pages}{3865} (\bibinfo{year}{1996}).

\bibitem[{lsd()}]{lsdanote}
\bibinfo{note}{The calculations were repeated using a local spin density
  approximation [W. Perdew and Y. Wang, Phys. Rev. B {\bf 45}, 13244 (1992)].
  No significant differerences were found.}

\bibitem[{\citenamefont{Grybos et~al.}(1994)\citenamefont{Grybos, Hohlwein,
  Zeiske, Sonntag, Kubanek, Eichhorn, and Wolf}}]{GrybosHZSKEW94}
\bibinfo{author}{\bibfnamefont{J.}~\bibnamefont{Grybos}},
  \bibinfo{author}{\bibfnamefont{D.}~\bibnamefont{Hohlwein}},
  \bibinfo{author}{\bibfnamefont{T.}~\bibnamefont{Zeiske}},
  \bibinfo{author}{\bibfnamefont{R.}~\bibnamefont{Sonntag}},
  \bibinfo{author}{\bibfnamefont{F.}~\bibnamefont{Kubanek}},
  \bibinfo{author}{\bibfnamefont{K.}~\bibnamefont{Eichhorn}}, \bibnamefont{and}
  \bibinfo{author}{\bibfnamefont{T.}~\bibnamefont{Wolf}},
  \bibinfo{journal}{Physica C} \textbf{\bibinfo{volume}{220}},
  \bibinfo{pages}{138} (\bibinfo{year}{1994}).

\bibitem[{\citenamefont{Grybos et~al.}(2001)\citenamefont{Grybos, Wabia,
  Guskos, and Typek}}]{GrybosWGT01}
\bibinfo{author}{\bibfnamefont{J.}~\bibnamefont{Grybos}},
  \bibinfo{author}{\bibfnamefont{M.}~\bibnamefont{Wabia}},
  \bibinfo{author}{\bibfnamefont{N.}~\bibnamefont{Guskos}}, \bibnamefont{and}
  \bibinfo{author}{\bibfnamefont{J.}~\bibnamefont{Typek}},
  \bibinfo{journal}{Molecular Physics Reports} \textbf{\bibinfo{volume}{34}},
  \bibinfo{pages}{121} (\bibinfo{year}{2001}).

\bibitem[{\citenamefont{Mazin et~al.}(1992)\citenamefont{Mazin, Jepsen,
  Andersen, Liechtenstein, Rashkeev, and Uspenskii}}]{MazinJALRU92}
\bibinfo{author}{\bibfnamefont{I.~I.} \bibnamefont{Mazin}},
  \bibinfo{author}{\bibfnamefont{O.}~\bibnamefont{Jepsen}},
  \bibinfo{author}{\bibfnamefont{O.~K.} \bibnamefont{Andersen}},
  \bibinfo{author}{\bibfnamefont{A.~I.} \bibnamefont{Liechtenstein}},
  \bibinfo{author}{\bibfnamefont{S.~N.} \bibnamefont{Rashkeev}},
  \bibnamefont{and} \bibinfo{author}{\bibfnamefont{Y.~A.}
  \bibnamefont{Uspenskii}}, \bibinfo{journal}{Phys. Rev. B}
  \textbf{\bibinfo{volume}{45}}, \bibinfo{pages}{5103} (\bibinfo{year}{1992}).

\bibitem[{\citenamefont{Schabel et~al.}(1998)\citenamefont{Schabel, Park,
  Matsuura, Shen, Bonn, Liang, and Hardy}}]{SchabelMPMSBLRH98}
\bibinfo{author}{\bibfnamefont{M.~C.} \bibnamefont{Schabel}},
  \bibinfo{author}{\bibfnamefont{C.-H.} \bibnamefont{Park}},
  \bibinfo{author}{\bibfnamefont{A.}~\bibnamefont{Matsuura}},
  \bibinfo{author}{\bibfnamefont{Z.-X.} \bibnamefont{Shen}},
  \bibinfo{author}{\bibfnamefont{D.~A.} \bibnamefont{Bonn}},
  \bibinfo{author}{\bibfnamefont{R.}~\bibnamefont{Liang}}, \bibnamefont{and}
  \bibinfo{author}{\bibfnamefont{W.~N.} \bibnamefont{Hardy}},
  \bibinfo{journal}{Phys. Rev. B} \textbf{\bibinfo{volume}{57}},
  \bibinfo{pages}{6090} (\bibinfo{year}{1998}).

\bibitem[{\citenamefont{Mackenzie et~al.}(1996)\citenamefont{Mackenzie, Julian,
  Diver, Mcmullan, Ray, Lonzarich, Maeno, Nishizaki, and
  Fujita}}]{MackenzieJDMRLMNF96}
\bibinfo{author}{\bibfnamefont{A.~P.} \bibnamefont{Mackenzie}},
  \bibinfo{author}{\bibfnamefont{S.~R.} \bibnamefont{Julian}},
  \bibinfo{author}{\bibfnamefont{A.~J.} \bibnamefont{Diver}},
  \bibinfo{author}{\bibfnamefont{G.~J.} \bibnamefont{Mcmullan}},
  \bibinfo{author}{\bibfnamefont{M.~P.} \bibnamefont{Ray}},
  \bibinfo{author}{\bibfnamefont{G.~G.} \bibnamefont{Lonzarich}},
  \bibinfo{author}{\bibfnamefont{Y.}~\bibnamefont{Maeno}},
  \bibinfo{author}{\bibfnamefont{S.}~\bibnamefont{Nishizaki}},
  \bibnamefont{and} \bibinfo{author}{\bibfnamefont{T.}~\bibnamefont{Fujita}},
  \bibinfo{journal}{Phys. Rev. Lett.} \textbf{\bibinfo{volume}{76}},
  \bibinfo{pages}{3786} (\bibinfo{year}{1996}).

\bibitem[{\citenamefont{Carrington et~al.}(2003)\citenamefont{Carrington,
  Meeson, Cooper, Balicas, Hussey, Yelland, Lee, Yamamoto, Tajima, Kazakov
  et~al.}}]{CarringtonMCBHYLYTKK03}
\bibinfo{author}{\bibfnamefont{A.}~\bibnamefont{Carrington}},
  \bibinfo{author}{\bibfnamefont{P.~J.} \bibnamefont{Meeson}},
  \bibinfo{author}{\bibfnamefont{J.~R.} \bibnamefont{Cooper}},
  \bibinfo{author}{\bibfnamefont{L.}~\bibnamefont{Balicas}},
  \bibinfo{author}{\bibfnamefont{N.~E.} \bibnamefont{Hussey}},
  \bibinfo{author}{\bibfnamefont{E.~A.} \bibnamefont{Yelland}},
  \bibinfo{author}{\bibfnamefont{S.}~\bibnamefont{Lee}},
  \bibinfo{author}{\bibfnamefont{A.}~\bibnamefont{Yamamoto}},
  \bibinfo{author}{\bibfnamefont{S.}~\bibnamefont{Tajima}},
  \bibinfo{author}{\bibfnamefont{S.~M.} \bibnamefont{Kazakov}},
  \bibnamefont{et~al.}, \bibinfo{journal}{Phys. Rev. Lett.}
  \textbf{\bibinfo{volume}{91}}, \bibinfo{pages}{037003}
  (\bibinfo{year}{2003}).

\bibitem[{\citenamefont{Carrington et~al.}(2007)\citenamefont{Carrington,
  Yelland, Fletcher, and Cooper}}]{CarringtonYFC07}
\bibinfo{author}{\bibfnamefont{A.}~\bibnamefont{Carrington}},
  \bibinfo{author}{\bibfnamefont{E.~A.} \bibnamefont{Yelland}},
  \bibinfo{author}{\bibfnamefont{J.~D.} \bibnamefont{Fletcher}},
  \bibnamefont{and} \bibinfo{author}{\bibfnamefont{J.~R.}
  \bibnamefont{Cooper}}, \bibinfo{journal}{Physica C}
  \textbf{\bibinfo{volume}{456}}, \bibinfo{pages}{92} (\bibinfo{year}{2007}).

\bibitem{ElfimovSD07}I.S. Elfimov, G.A. Sawatzky, and A. Damascelli,
  arXiv:0706.4276v1 [cond-mat.str-el].

\bibitem[{\citenamefont{Ambroschdraxl et~al.}(1991)\citenamefont{Ambroschdraxl,
  Blaha, and Schwarz}}]{AmbroschdraxlBS91}
\bibinfo{author}{\bibfnamefont{C.}~\bibnamefont{Ambroschdraxl}},
  \bibinfo{author}{\bibfnamefont{P.}~\bibnamefont{Blaha}}, \bibnamefont{and}
  \bibinfo{author}{\bibfnamefont{K.}~\bibnamefont{Schwarz}},
  \bibinfo{journal}{Phys. Rev. B} \textbf{\bibinfo{volume}{44}},
  \bibinfo{pages}{5141} (\bibinfo{year}{1991}).

\bibitem[{\citenamefont{Lanzara et~al.}(2001)\citenamefont{Lanzara, Bogdanov,
  Zhou, Kellar, Feng, Lu, Yoshida, Eisaki, Fujimori, Kishio
  et~al.}}]{LanzaraBZKFLYEFKSNUHS01}
\bibinfo{author}{\bibfnamefont{A.}~\bibnamefont{Lanzara}},
  \bibinfo{author}{\bibfnamefont{P.~V.} \bibnamefont{Bogdanov}},
  \bibinfo{author}{\bibfnamefont{X.~J.} \bibnamefont{Zhou}},
  \bibinfo{author}{\bibfnamefont{S.~A.} \bibnamefont{Kellar}},
  \bibinfo{author}{\bibfnamefont{D.~L.} \bibnamefont{Feng}},
  \bibinfo{author}{\bibfnamefont{E.~D.} \bibnamefont{Lu}},
  \bibinfo{author}{\bibfnamefont{T.}~\bibnamefont{Yoshida}},
  \bibinfo{author}{\bibfnamefont{H.}~\bibnamefont{Eisaki}},
  \bibinfo{author}{\bibfnamefont{A.}~\bibnamefont{Fujimori}},
  \bibinfo{author}{\bibfnamefont{K.}~\bibnamefont{Kishio}},
  \bibnamefont{et~al.}, \bibinfo{journal}{Nature}
  \textbf{\bibinfo{volume}{412}}, \bibinfo{pages}{510} (\bibinfo{year}{2001}).

\bibitem[{\citenamefont{Yu et~al.}(1991)\citenamefont{Yu, Park, and
  Freeman}}]{YuPF91}
\bibinfo{author}{\bibfnamefont{J.~J.} \bibnamefont{Yu}},
  \bibinfo{author}{\bibfnamefont{K.~T.} \bibnamefont{Park}}, \bibnamefont{and}
  \bibinfo{author}{\bibfnamefont{A.~J.} \bibnamefont{Freeman}},
  \bibinfo{journal}{Physica C} \textbf{\bibinfo{volume}{172}},
  \bibinfo{pages}{467} (\bibinfo{year}{1991}).

\end{thebibliography}
\end{document}